%% file: main.tex
\documentclass[letterpaper,twocolumn,10pt]{article}
\usepackage{usenix,epsfig,endnotes}

\usepackage{epsfig,endnotes,algorithm,verbatim,algpseudocode}
\usepackage{amsmath}
\usepackage{balance}
\usepackage{flushend}
\usepackage{multirow}
\usepackage{color}
\usepackage{wrapfig}
\usepackage{graphicx}
\usepackage{url}
\usepackage{subcaption}
\usepackage{booktabs}
\usepackage{mwe}
\usepackage{hyperref}
\usepackage{times}
\usepackage{datetime}
\usepackage{url}
\usepackage{array}
\usepackage[numbers]{natbib}
\usepackage{enumitem}
\usepackage[utf8]{inputenc}
\usepackage[T1]{fontenc}

\newcommand{\todo}[1]{\textcolor{red}{#1}}
\newcommand{\name}[0]{Bené}

\begin{document}

\date{}

\title{\Large \bf \name: On Demand Cost-Effective Scaling at the Edge}

\author{
 {\rm Faria Kalim}\\
 University of Illinois, Urbana Champaign
 \\
 kalim2@illinois.edu
 \and
 {\rm Shadi A. Noghabi}\\
	University of Illinois, Urbana Champaign
	\\
	abdolla2@illinois.edu
} 

\maketitle

\thispagestyle{empty}

\input{tex/abstract.tex}
\input{tex/introduction}
\input{tex/design}
\input{tex/related-work.tex}
\input{tex/future-work.tex}
\input{tex/conclusion.tex}

{\footnotesize \bibliographystyle{acm}
\bibliography{bib/kalim2-hotedge18.bib}}


\end{document}

%% file: tex/abstract.tex
\subsection*{Abstract}

Edge computing has become increasingly popular across many domains and enterprises. However, given the locality constraint of edges (i.e., only close-by edges are useful), multiplexing diverse workloads becomes challenging. This results in poor resource utilization in edge resources  that are provisioned for  peak demand. 
A simple way to allow multiplexing is through micro-data centers, that bring computation close to the users while supporting diverse workloads throughout the data, along with edges.
In this paper, we argue for a hybrid approach of dedicated edge resources within an enterprise and on demand resources in micro-data centers that are shared across entities. We show that this hybrid approach is an effective and cost-efficient way for scaling workloads and removes the need for over-provisioning dedicated resources per enterprise. Moreover, compared to a scaling approach that uses only the edges across enterprises, micro-data centers also form a trusted third party that can maintain important quality of service guarantees such as data privacy, security, and availability.


%% file: tex/introduction.tex
\section{Introduction}
\label{sec:intro}

\begin{figure*}[t]
\includegraphics[width=\textwidth]{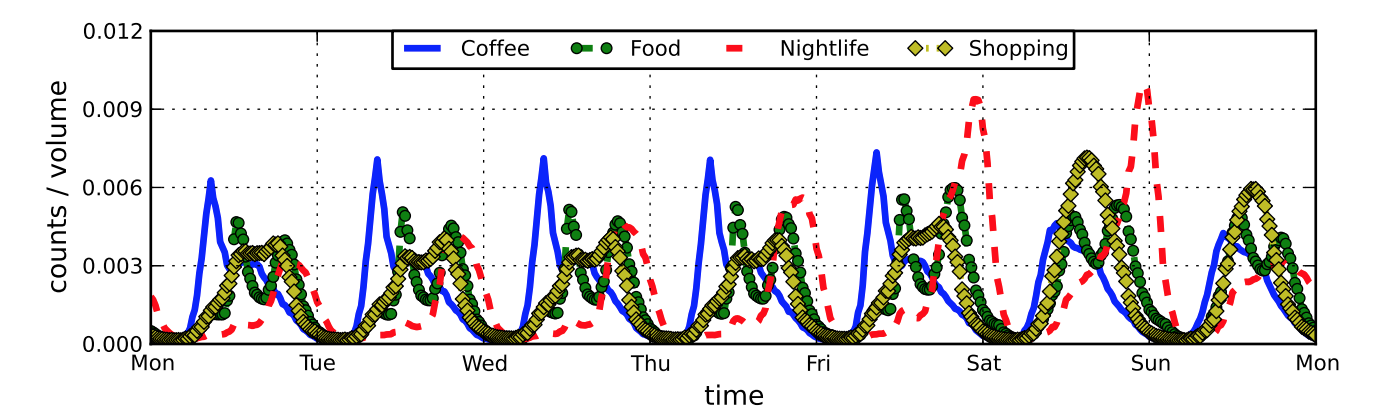}
\caption{ Weekly patterns of four broad categories of real-world activities as reflected in Foursquare checkins. Taken from~\cite{grinberg2013extracting}}
\label{diurnal-patterns}
\centering
\end{figure*}

\begin{figure}[t]
\includegraphics[width=\columnwidth]{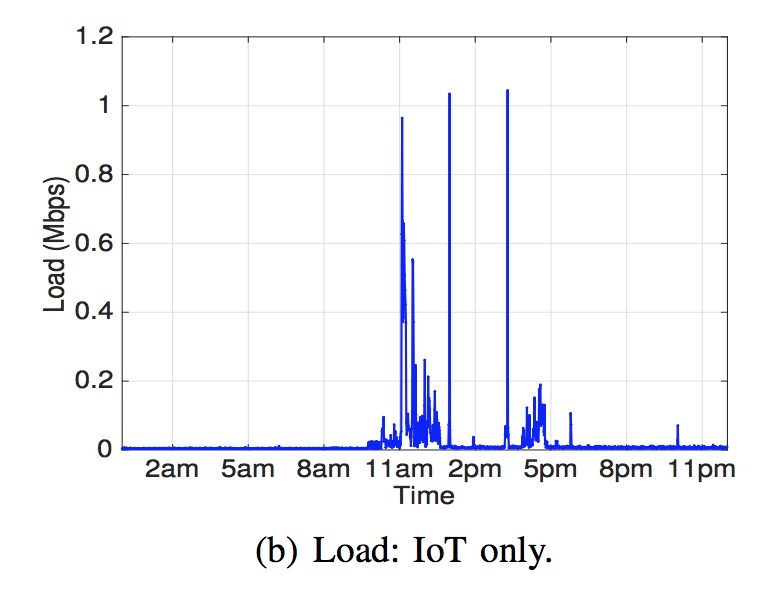}
\caption{ Network load from IoT devices on a representative day in a smart-campus environment. Taken from~\cite{sivanathan2017characterizing}}
\label{peak}
\centering
\end{figure}

Edge computing---which pushes computation from the cloud closer to the end user---has recently piqued significant interest. The paradigm has several advantages including reducing end-to-end latency, providing continuous service despite intermittent connectivity to the cloud, reducing network bandwidth consumption, and reducing the monetary cost of using cloud services~\cite{ fog-iot, edge-driver, edge-vision-2,nebula, satya-edge,   satya-hostile-env, cloudlet-bahl, edge-vision}. As a result, the edge has become a good fit for a wide variety of domains, such as the Internet of Things (IoT), video processing, online gaming, or augmented/virtual reality~\cite{ananthanarayanan2017real, fog-iot, edge-vision, farmbeats, yi2015survey,  zhu2013improving}.

Most application workloads are dynamic with diurnal, weekly, seasonal (e.g., holidays), or occasional (e.g., sports event) peaks and drops. As an example, Figure~\ref{diurnal-patterns} shows workload patterns from four real-world activities (coffee purchases, food purchases, nightlife and shopping). The figure is taken from~\cite{grinberg2013extracting}.  As another example, Figure~\ref{peak} shows the network traffic from IoT devices on a smart college campus on a single representative day. This figure is 
taken from~\cite{sivanathan2017characterizing}.
Dynamic patterns are particularly more severe at the edge given its locality constraints. Edges only serve applications within a vicinity. Thus, they lose the luxury of aggregating data globally (in a single data-center) and potentially masking some of the variation.

Many of these patterns are predictable, enabling application developers to plan ahead and scale accordingly. However, due to limited application diversity within an enterprise, the potential for multiplexing will be poor.  For instance, a university and the traffic monitoring system in a city may each have their own set of edges. Since each entity will have a relatively smaller number of edges (compared to a data-center), and less diversity among its workloads, e.g., the peak for most applications in a university is 9am to 5pm, and traffic related application are at rush hour, the potential for multiplexing \emph{within} each enterprise is poor. 
Therefore, developers have a choice---either to observe degraded performance at peak times or  over-provision resources for the peak (which wastes resource during low load). While the former choice is untenable, the latter can be expensive.

We argue for the need of \emph{a hybrid approach of dedicated resources within an enterprise and on demand resources in micro-data centers (MDCs) that are shared across entities.} This hybrid approach brings the benefits of multiplexing, as on the cloud, closer to the edge. Along with scalability, it also enables fast testing and development of edge applications without needing to provision new edges, and allows fault tolerance until new edge devices can be permanently added (which can be a lengthy process). 

However, providing on demand scaling at the edge has many challenges and complexities due to the nature of the  new environment. Dedicated over-provisioned resources provide the guarantee that not only are there  resources available at the peak, but that they are also in an acceptable \textit{proximity and locality} of the users. 
An on-demand scalability approach through MDCs should be able to guarantee resources while satisfying locality constraints.  To do so, resources should be provisioned ahead of time in a locality and in a demand-aware fashion. 

Moreover, on demand scaling should also be cost-effective for both the enterprises using it and the MDC provider. The provider takes advantage of the fact that even though an enterprise individually has poor multiplexing,  combining workloads with diverse peaks from multiple entities provides efficient multiplexing. Conversely, enterprises require certain guarantees such as availability, which translates to dedicating resources at the provider. Although stringent guarantees entail higher costs, having a pricing model that can motivate more entities to use MDCs and in turn, increase diversity and utilization is crucial. The benefit of on demand scaling at peak loads should outweigh buying individual resources, but should not come at a loss for providers. 

In this work, we propose {\name}, a shared MDC infrastructure that provides on demand scalability on the edge. {\name} consists of a shared pool of dedicated resources, that can be used across multiple enterprises. {\name} manages the pool of resources end-to-end from capacity planning, constructing pricing models, to scheduling resources for incoming demands and actually assigning and monitoring resources. {\name} plans the capacity of each MDC in a location-aware manner, using the historical demands of each location.

{\name} provides two levels of resource guarantees: a) \textit{reserved allocations} which are guaranteed resources similar to  leased resources for a predefined time-frame with a fixed price, e.g., 2 units at 4pm -- 6pm everyday, and b) \textit{real-time allocations} which are allocated in real-time based on current availability and dynamic pricing based on demand. Reserved resources provide similar guarantees to dedicated resources and real-time resources help increase overall utilization of the system, making it a cost-effective solution for the provider.   

The contribution of this paper includes describing an infrastructure based on MDCs that can be used to scale the edge cheaply and on-demand. Section~\ref{sec:design} describes the high-level design of {\name}, a simple pricing model, and its scheduler. Section~\ref{sec:related-work} discusses related work in the area and Section ~\ref{sec:future-work} describes possible future directions. We conclude with Section~\ref{sec:conclusion}.

%% file: tex/design.tex
\section{Design}
\label{sec:design}

\begin{figure}[tp]
\includegraphics[width=\columnwidth]{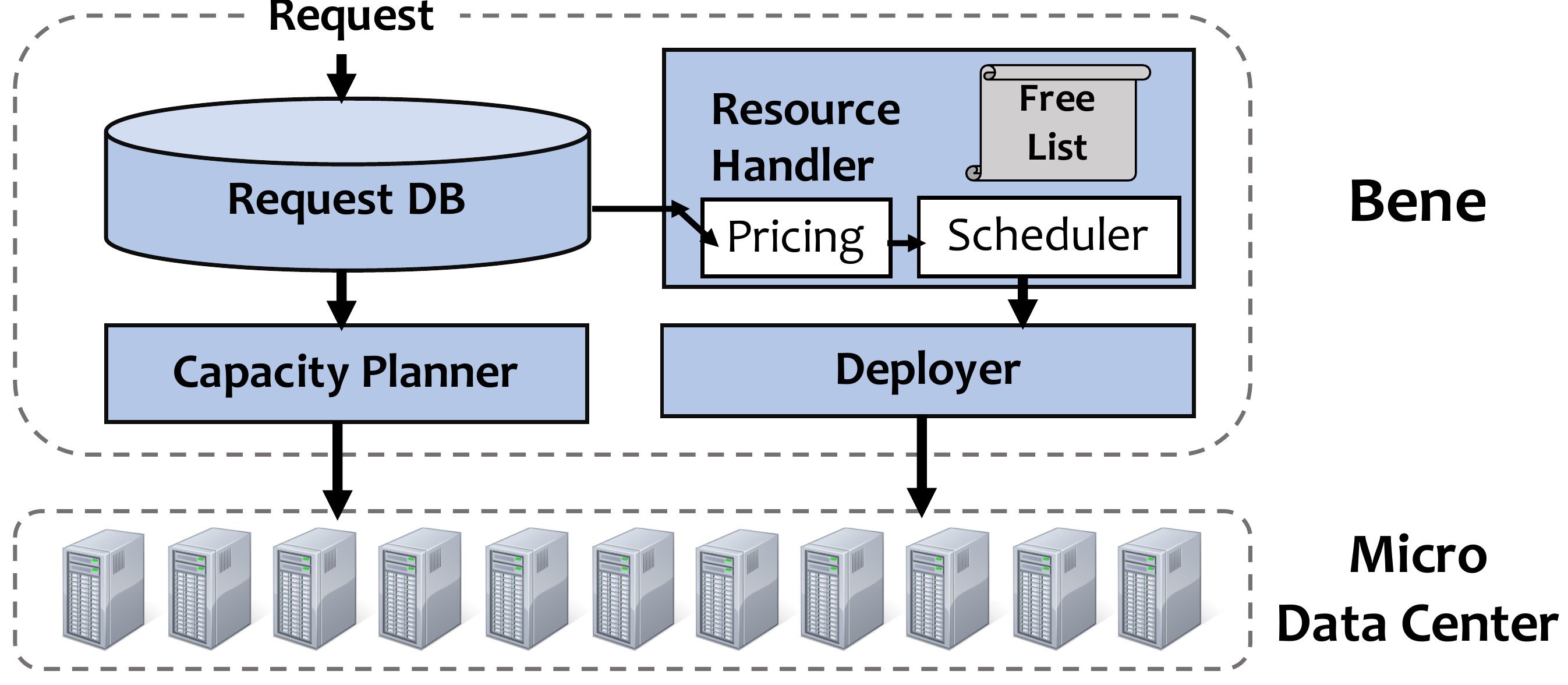}
\caption{{\name} Architecture}
\label{architecture}
\centering
\end{figure}

Figure~\ref{architecture} shows the high level pipeline that comprises {\name}. As resource requests come in from enterprises, they are stored in a database and can be later used for capacity planning for the edge (Section~\ref{capacity-planning}).  More urgently, the resource handler (Section~\ref{resource-provisioning}) uses a pricing policy to decide the amount to be charged for serving the request, and passes it on to the scheduler that comes up with a scheduling plan for the request. Finally, the deployer (Section ~\ref{deployer}) reserves resources on the MDC according to the scheduling plan at the requested time, provisions containers, and starts the job.

Figure~\ref{architecture} shows a single instance of {\name} in a region. Using MDCs to scale allows the architecture to be decentralized: as there is no dependency between regions, a similar MDC can be placed in any region that requires scalability on the edge. The largest possible region size that {\name} can handle is determined by factors such as population and load density in an area, number of enterprises that require service, the number of edge machines that need to be set up and management costs and is as such, out of the scope of this work. 

\subsection{Capacity Planning}
\label{capacity-planning}

{\name} measures how load changes over time in a particular region to determine the amount of resources that an MDC should be provisioned with. The request database maintains information about historical requests which can be used to predict the amount of resources required to adapt to the changing load. This information includes 1) the number of instances required per request, 2) the date and time for which they are leased, and 3) a service level objective in terms of latency and throughput. The database also records which requests {\name} was able to service and which requests it was not able to service because of a lack of available resources. This allows Bene to estimate the individual peak loads, simultaneous cumulative peak loads from multiple enterprises, the expected service level objectives (SLOs) and the actual performance of the edges. Then, extra machines can be provisioned accordingly to adapt to the extra load.

Capacity planning can be done periodically, e.g., before major events. Additionally, a good trigger for capacity planning is when an increasingly large number of requests are unable to meet their required service level objectives or extensive use of manual administrative control (bypassing Bene) .

\subsection{Resource Provisioning}
\label{resource-provisioning}

The resource provisioning module is responsible for calculating a price for a user's resource request and for developing a scheduling plan to service the request.

{\name} allows two kinds of requests: \textit{reservations} and \textit{real-time} requests. Both requests contain the number of instances required, the specifications of the instances in terms of resources, the time frame for the request, and the expected latency and throughput service level objectives. Each instance requested can be a VM/container and we will refer to them as slots from now on. We assume that a user does not specify objectives that are unreasonable: for example, a system may never be able to support a latency of less than 1 ms. 

Reservation requests can be taught of as a lease on the resource. Reservations are made in advance by an enterprise, which increases the chances of resources being available so the request can be serviced. The request must come sufficiently in advance (currently set to 15 min in {\name}) for it to be considered a reservation. This time is necessary to make scheduling decisions and prepare slots to start jobs on time. {\name} maintains a free list of resources, for every time unit (15-min intervals in practice). For each time unit, the free list records which slots in the MDC are available for use. When a reservation comes in, {\name} uses the free list to assign the request appropriate slots based on the given requirements and then updates the free list. 

Real-time requests are allowed in {\name} to increase resource utilization (for the provider) and utilize the unused time slots. However, real-time request are scheduled in a best-effort manner. Customers run the risk that no resources may be available to service real-time requests. 

\paragraph{Pricing:} Bene's pricing structure is similar to pricing structures in the cloud with dedicated resources and spot instance \cite{ec2-pricing,spot-pricing}. {\name} settles on a defined price per slot based on the amount of the resources it provides and the number of them that are available. These defined prices are used to calculate how much the user is charged for a reservation. Similar to the pricing of dedicated instances in AWS, these instances have higher prices as they are guaranteed to be available to the user for the entire duration of their job.

However, {\name} reduces the prices of instances that are unused in real-time, and uses these prices to charge real-time requests. This increases revenue as it encourages customers to make more real-time requests. However, before a real-time request is scheduled, the customer is told that the resources might be taken away at a given time because they have been scheduled for reserved requests.

Thus, the price that a customer is charged can be calculated by:\\

$Charge = Price\ Rate \times Number\ of\ Containers \times Container\ Type \times Occupancy\ Duration$

\paragraph{Scheduling:}

The scheduler runs periodically (every 15 mins), makes a note of the jobs that must be stopped in the next time interval and creates a schedule for the jobs that have to be started. 

When requests come in, the request handler ensures that slots of the required kind are available for the requested duration. However, the scheduler ensures that slots are scheduled in the most efficient way, i.e., its goal is to pack slots onto machines to minimize fragmentation and increase resource utilization. To be able to do so, {\name} uses the scheduling algorithm from Rayon~\cite{curino2014reservation}. Rayon is a good choice for the system because it allows us to change the current resource reservation on the cluster quickly and still create a close to optimal solution.

\subsection{Deployer}
\label{deployer}

The deployer receives a schedule and accordingly, closes down the containers that are at the end of their jobs, cleans up state, ships the binaries for the new jobs to their respective slots, and starts containers there. This allows new jobs to start exactly on time. Customers are also allowed access to their containers before the starting time so they can update their applications with the correct endpoints.

The deployer's most important job is to detect failures in the MDC, and report the failure to the scheduler. As an MDC is relatively small, the deployer adopts a simple approach of pinging the slots  at regular intervals to make sure that they are available. Once a failure is detected, the deployer informs the scheduler and attempts to restart the slot. If the attempt fails, then it fetches the new schedule to re-provision a container for the job. The pricing module is updated accordingly about the loss of service for the given duration. It is possible that no containers might be available to re-provision the job, in which case, a slot for a real-time request can be pre-empted. If there are no real-time requests running on the cluster, then the customer is informed accordingly and the price charged is updated.

%% file: tex/related-work.tex
\section{Related Work} 
\label{sec:related-work}

\subsection{Quality of Service at the Edge}
Quality of service metrics such as availability and performance are crucial to meet the low-latency service level objectives of the applications running  on the edge. ~\cite{vaquero2014finding} surveys edge computing and claims that availability on the edge is highly variable, an attribute inherited from peer-to-peer architectures. ~\cite{madsen2013reliability} surveys existing paradigms including cloud computing and concludes analytically that building a reliable edge computing environment is difficult but feasible, but does not propose solutions to any of the challenges described.

In terms of performance, ~\cite{narayanan2014towards}  describes a simple analytical model to measure the amount of excess capacity in a given edge deployment that is able to meet a given latency threshold while sustaining the probability of a single failure. Several works exist that use the edge to improve the performance of their applications e.g. Cachier~\cite{drolia2017cachier} is a system that uses the edge as an efficient cache for image recognition applications and uses dynamic load balancing  between the edge and the cloud to provide the low latency responses. Similarly, ~\cite{ananthanarayanan2017real} describes how video analytics is the killer app for the edge, as the paradigm is very effective for returning low latency responses, while using little bandwidth. ~\cite{deng2016optimal} describes the tradeoff between power consumption and transmission delay in a geo-distributed deployment. The work formulates and solves a workload allocation problem which suggests the optimal workload allocations between fog and cloud towards the minimal power consumption with the constrained service delay. 

\subsection{Resource Allocation on the Edge} Related work has considered the problem of task allocation to different kinds of edge deployments. iFogSim ~\cite{gupta2017ifogsim} is a simulator that models IoT and fog environments and can measure the impact of resource management techniques in terms of latency, network congestion, energy consumption, and cost. ~\cite{yao2017heterogeneous} solves a linear programming, and a faster low-complexity heuristic to find a good allocation of jobs onto heterogeneous machines in different cloudlets. ~\cite{gomes2016software} uses software-defined networking to manage accessibility to edge resources, while addressing network events such as traffic overload and failures that can affect quality of service. ~\cite{aazam2015dynamic} proposes a framework that considers the problem of high user churn rate and manages resources accordingly.

\subsection{Scalability in other paradigms}
As much as scalability is essential for supporting the workloads of our times, it is not easy to achieve in practice. Even though a data center allows cheaper multiplexing, it is difficult to maximizing its utilization at all times~\cite{greenberg2008cost}. A large body of work has considered the problem of applications that can scale to several thousand machines e.g.~\cite{aguilera2007sinfonia,rowstron2001pastry,abadi2016tensorflow,weil2006ceph,hendrickson2016serverless}. An essential problem is partitioning state efficiently, and if required, keeping it synchronized with other replicated instances: solutions range from efficient state partitioning ~\cite{agrawal2010data} to expressing computation in a way that does not require synchronization~\cite{dean2008mapreduce}.

%% file: tex/future-work.tex
\section{Discussion \& Future Work}
\label{sec:future-work}

\textbf{MDCs as opposed to other approaches: } Alternate solutions to MDCs include building shared resource pools   of unused resources owned by individual enterprises. This allows enterprises to purchase specialized dedicated resources. However, a few problems can pose hurdles in forming a platform from these resources. For example, enterprises can be reluctant about sending their data to another enterprise for privacy reasons. In addition, the application would have to support re-routing of requests: if a resource fails at an enterprise and another resource is provisioned on the premises of another enterprises, users have to be informed that they must now contact the other enterprise. Fault tolerance through overprovisioning is significantly easier to do within an MDC.\\

\noindent\textbf{Capacity Planning:} Setting up an MDC allows load to be shared cheaply but comes with a large initial investment. Administrators can optimize the number and capacities of MDCs to minimize the initial cost while trying to provide service in as large an area as possible. The first aspect to consider is the number of MDCs in a region: while it is cheaper to set up a single MDC, having multiple MDCs provides availability in the face of large-scale failures due to natural disasters or accidents. If failures are unlikely, a single MDC can be responsible for a single region.

Additionally,   the size of a region that an MDC can cater to is worth exploring. This depends on several factors such as the population density in the region, the workload experienced, the stringency of the service level objectives and the extent to which the MDC can be scaled up or out. \\

\noindent\textbf{Resource Provisioning:} Setting up dedicated edges can be enticing for enterprises that have very specific resource demands. For example, they may require specialized hardware for machine learning or applications of crypotcurrency. It would be useful to perform a survey of the kinds of resources users require and provision accordingly. Providing specialized resources of course comes at the cost of less opportunity for multiplexing in case other users do not have the same resource demands, and less revenue for {\name}.

%% file: tex/conclusion.tex
\section{Conclusion}
\label{sec:conclusion}

In this paper, we showed that a hybrid approach of dedicated edges and on-demand micro-data centers form the most viable solution for scaling in a cost-effective manner at the edge, while preserving important quality of service properties such as security and availability.

%% file: main.bbl
\begin{thebibliography}{10}

\bibitem{ec2-pricing}
Amazon ec2 pricing.
\newblock \url{https://aws.amazon.com/ec2/pricing/}.
\newblock Accessed: \today.

\bibitem{spot-pricing}
Amazon ec2 spot instances pricing.
\newblock \url{https://aws.amazon.com/ec2/spot/pricing/}.
\newblock Accessed: \today.

\bibitem{aazam2015dynamic}
{\sc Aazam, M., and Huh, E.-N.}
\newblock Dynamic resource provisioning through fog micro datacenter.
\newblock In {\em Pervasive Computing and Communication Workshops (PerCom
  Workshops), 2015 IEEE International Conference on\/} (2015), IEEE,
  pp.~105--110.

\bibitem{abadi2016tensorflow}
{\sc Abadi, M., Agarwal, A., Barham, P., Brevdo, E., Chen, Z., Citro, C.,
  Corrado, G.~S., Davis, A., Dean, J., Devin, M., et~al.}
\newblock Tensorflow: Large-scale machine learning on heterogeneous distributed
  systems.
\newblock {\em arXiv preprint arXiv:1603.04467\/} (2016).

\bibitem{agrawal2010data}
{\sc Agrawal, D., El~Abbadi, A., Antony, S., and Das, S.}
\newblock Data management challenges in cloud computing infrastructures.
\newblock In {\em International Workshop on Databases in Networked Information
  Systems\/} (2010), Springer, pp.~1--10.

\bibitem{aguilera2007sinfonia}
{\sc Aguilera, M.~K., Merchant, A., Shah, M., Veitch, A., and Karamanolis, C.}
\newblock Sinfonia: a new paradigm for building scalable distributed systems.
\newblock In {\em ACM SIGOPS Operating Systems Review\/} (2007), vol.~41, ACM,
  pp.~159--174.

\bibitem{ananthanarayanan2017real}
{\sc Ananthanarayanan, G., Bahl, P., Bod{\'\i}k, P., Chintalapudi, K.,
  Philipose, M., Ravindranath, L., and Sinha, S.}
\newblock Real-time video analytics: The killer app for edge computing.
\newblock {\em Computer 50}, 10 (2017), 58--67.

\bibitem{fog-iot}
{\sc Bonomi, F., Milito, R., Zhu, J., and Addepalli, S.}
\newblock Fog computing and its role in the internet of things.
\newblock In {\em Proceedings of the first edition of the MCC workshop on
  Mobile cloud computing\/} (2012), ACM, pp.~13--16.

\bibitem{edge-driver}
{\sc Carlini, S.}
\newblock The drivers and benefits of edge computing.
\newblock {\em Schneider Electric--Data Center Science Center\/} (2016), 8.

\bibitem{curino2014reservation}
{\sc Curino, C., Difallah, D.~E., Douglas, C., Krishnan, S., Ramakrishnan, R.,
  and Rao, S.}
\newblock Reservation-based scheduling: If you're late don't blame us!
\newblock In {\em Proceedings of the ACM Symposium on Cloud Computing\/}
  (2014), ACM, pp.~1--14.

\bibitem{dean2008mapreduce}
{\sc Dean, J., and Ghemawat, S.}
\newblock Mapreduce: simplified data processing on large clusters.
\newblock {\em Communications of the ACM 51}, 1 (2008), 107--113.

\bibitem{deng2016optimal}
{\sc Deng, R., Lu, R., Lai, C., Luan, T.~H., and Liang, H.}
\newblock Optimal workload allocation in fog-cloud computing toward balanced
  delay and power consumption.
\newblock {\em IEEE Internet of Things Journal 3}, 6 (2016), 1171--1181.

\bibitem{drolia2017cachier}
{\sc Drolia, U., Guo, K., Tan, J., Gandhi, R., and Narasimhan, P.}
\newblock Cachier: Edge-caching for recognition applications.
\newblock In {\em Distributed Computing Systems (ICDCS), 2017 IEEE 37th
  International Conference on\/} (2017), IEEE, pp.~276--286.

\bibitem{edge-vision-2}
{\sc Garcia~Lopez, P., Montresor, A., Epema, D., Datta, A., Higashino, T.,
  Iamnitchi, A., Barcellos, M., Felber, P., and Riviere, E.}
\newblock Edge-centric computing: Vision and challenges.
\newblock {\em ACM SIGCOMM Computer Communication Review 45}, 5 (2015), 37--42.

\bibitem{gomes2016software}
{\sc Gomes, R.~L., Bittencourt, L.~F., Madeira, E.~R., Cerqueira, E.~C., and
  Gerla, M.}
\newblock Software-defined management of edge as a service networks.
\newblock {\em IEEE Transactions on Network and Service Management 13}, 2
  (2016), 226--239.

\bibitem{greenberg2008cost}
{\sc Greenberg, A., Hamilton, J., Maltz, D.~A., and Patel, P.}
\newblock The cost of a cloud: research problems in data center networks.
\newblock {\em ACM SIGCOMM computer communication review 39}, 1 (2008), 68--73.

\bibitem{grinberg2013extracting}
{\sc Grinberg, N., Naaman, M., Shaw, B., and Lotan, G.}
\newblock Extracting diurnal patterns of real world activity from social media.
\newblock In {\em ICWSM\/} (2013).

\bibitem{gupta2017ifogsim}
{\sc Gupta, H., Vahid~Dastjerdi, A., Ghosh, S.~K., and Buyya, R.}
\newblock ifogsim: A toolkit for modeling and simulation of resource management
  techniques in the internet of things, edge and fog computing environments.
\newblock {\em Software: Practice and Experience 47}, 9 (2017), 1275--1296.

\bibitem{hendrickson2016serverless}
{\sc Hendrickson, S., Sturdevant, S., Harter, T., Venkataramani, V.,
  Arpaci-Dusseau, A.~C., and Arpaci-Dusseau, R.~H.}
\newblock Serverless computation with openlambda.
\newblock {\em Elastic 60\/} (2016), 80.

\bibitem{madsen2013reliability}
{\sc Madsen, H., Burtschy, B., Albeanu, G., and Popentiu-Vladicescu, F.}
\newblock Reliability in the utility computing era: Towards reliable fog
  computing.
\newblock In {\em Systems, Signals and Image Processing (IWSSIP), 2013 20th
  International Conference on\/} (2013), IEEE, pp.~43--46.

\bibitem{narayanan2014towards}
{\sc Narayanan, I., Kansal, A., Sivasubramaniam, A., Urgaonkar, B., and
  Govindan, S.}
\newblock Towards a leaner geo-distributed cloud infrastructure.
\newblock In {\em HotCloud\/} (2014).

\bibitem{rowstron2001pastry}
{\sc Rowstron, A., and Druschel, P.}
\newblock Pastry: Scalable, decentralized object location, and routing for
  large-scale peer-to-peer systems.
\newblock In {\em IFIP/ACM International Conference on Distributed Systems
  Platforms and Open Distributed Processing\/} (2001), Springer, pp.~329--350.

\bibitem{nebula}
{\sc Ryden, M., Oh, K., Chandra, A., and Weissman, J.}
\newblock Nebula: Distributed edge cloud for data intensive computing.
\newblock In {\em Cloud Engineering (IC2E), 2014 IEEE International Conference
  on\/} (2014), IEEE, pp.~57--66.

\bibitem{satya-edge}
{\sc Satyanarayanan, M.}
\newblock The emergence of edge computing.
\newblock {\em Computer 50}, 1 (2017), 30--39.

\bibitem{cloudlet-bahl}
{\sc Satyanarayanan, M., Bahl, P., Caceres, R., and Davies, N.}
\newblock The case for vm-based cloudlets in mobile computing.
\newblock {\em IEEE pervasive Computing 8}, 4 (2009).

\bibitem{satya-hostile-env}
{\sc Satyanarayanan, M., Lewis, G., Morris, E., Simanta, S., Boleng, J., and
  Ha, K.}
\newblock The role of cloudlets in hostile environments.
\newblock {\em IEEE Pervasive Computing 12}, 4 (2013), 40--49.

\bibitem{edge-vision}
{\sc Shi, W., Cao, J., Zhang, Q., Li, Y., and Xu, L.}
\newblock Edge computing: Vision and challenges.
\newblock {\em IEEE Internet of Things Journal 3}, 5 (2016), 637--646.

\bibitem{sivanathan2017characterizing}
{\sc Sivanathan, A., Sherratt, D., Gharakheili, H.~H., Radford, A., Wijenayake,
  C., Vishwanath, A., and Sivaraman, V.}
\newblock Characterizing and classifying iot traffic in smart cities and
  campuses.
\newblock In {\em Proc. IEEE INFOCOM Workshop SmartCity, Smart Cities Urban
  Comput.\/} (2017), pp.~1--6.

\bibitem{vaquero2014finding}
{\sc Vaquero, L.~M., and Rodero-Merino, L.}
\newblock Finding your way in the fog: Towards a comprehensive definition of
  fog computing.
\newblock {\em ACM SIGCOMM Computer Communication Review 44}, 5 (2014), 27--32.

\bibitem{farmbeats}
{\sc Vasisht, D., Kapetanovic, Z., Won, J., Jin, X., Chandra, R., Sinha, S.~N.,
  Kapoor, A., Sudarshan, M., and Stratman, S.}
\newblock {FarmBeats}: An iot platform for data-driven agriculture.
\newblock In {\em NSDI\/} (2017), pp.~515--529.

\bibitem{weil2006ceph}
{\sc Weil, S.~A., Brandt, S.~A., Miller, E.~L., Long, D.~D., and Maltzahn, C.}
\newblock Ceph: A scalable, high-performance distributed file system.
\newblock In {\em Proceedings of the 7th symposium on Operating systems design
  and implementation\/} (2006), USENIX Association, pp.~307--320.

\bibitem{yao2017heterogeneous}
{\sc Yao, H., Bai, C., Xiong, M., Zeng, D., and Fu, Z.}
\newblock Heterogeneous cloudlet deployment and user-cloudlet association
  toward cost effective fog computing.
\newblock {\em Concurrency and Computation: Practice and Experience 29}, 16
  (2017).

\bibitem{yi2015survey}
{\sc Yi, S., Li, C., and Li, Q.}
\newblock A survey of fog computing: concepts, applications and issues.
\newblock In {\em Proceedings of the 2015 Workshop on Mobile Big Data\/}
  (2015), ACM, pp.~37--42.

\bibitem{zhu2013improving}
{\sc Zhu, J., Chan, D.~S., Prabhu, M.~S., Natarajan, P., Hu, H., and Bonomi,
  F.}
\newblock Improving web sites performance using edge servers in fog computing
  architecture.
\newblock In {\em Service Oriented System Engineering (SOSE), 2013 IEEE 7th
  International Symposium on\/} (2013), IEEE, pp.~320--323.

\end{thebibliography}
